# Exact Safety Verification of Hybrid Systems Using Sums-Of-Squares Representation☆

Wang Lin[a,b], Min Wu[a], Zhengfeng Yang[a], Zhenbing Zeng[a]

[a] *Shanghai Key Laboratory of Trustworthy Computing, East China Normal University, Shanghai 200062, China*
[b] *College of Mathematics and Information Science, Wenzhou University, Zhejiang 325035, China*

## Abstract

In this paper we discuss how to generate inductive invariants for safety verification of hybrid systems. A hybrid symbolic-numeric method is presented to compute inequality inductive invariants of the given systems. A numerical invariant of the given system can be obtained by solving a parameterized polynomial optimization problem via sum-of-squares (SOS) relaxation. And a method based on Gauss-Newton refinement and rational vector recovery is deployed to obtain the invariants with rational coefficients, which *exactly* satisfy the conditions of invariants. Several examples are given to illustrate our algorithm.

*Keywords:* semidefinite programming, sum-of-squares relaxation, safety verification, invariant generation

## 1. Introduction

Complex physical systems are systems in which the techniques of sensing, control, communication and coordination are involved and interacted with

☆This material is supported in part by the Chinese National Natural Science Foundation under Grants 91018012(Wu, Yang, Zeng) and 10801052(Wu) and 10901055(Yang), and the Scientific Research Project of The Graduate School of East China Normal University under Grant CX2011009.

*Email addresses:* linwang@wzu.edu.cn (Wang Lin), mwu@sei.ecnu.edu.cn (Min Wu), zfyang@sei.ecnu.edu.cn (Zhengfeng Yang), zbzeng@sei.ecnu.edu.cn (Zhenbing Zeng)

each other. Among complex physical systems, many of them are safety critical systems, such as airplanes, railway, and automotive applications. Due to the complexity, ensuring correct functioning of these systems, e.g., spatial separation, especially collision avoidance of aircrafts during the entire flights, is among the most challenging and most important problems in computer science, mathematics and engineering.

As a common mathematical model for complex physical systems, hybrid systems [3, 8] are dynamical systems that are governed by interacting discrete and continuous dynamics [1, 6, 8]. Continuous dynamics is specified by differential equations, which is possibly subject to domain restrictions or algebraic relations resulting from physical circumstances or the interaction of continuous dynamics with discrete control. For discrete transitions, the hybrid system changes state instantaneously and possibly discontinuously, for example, the instantaneous change of control variables like the acceleration (e.g., the changing of $a$ by setting $a := -b$ with braking force $b > 0$).

The verification of hybrid systems is an important problem that has been studied extensively both by the control theory, and the formal verification community for over a decade. Among the most important verification questions for hybrid systems are those of *safety*, i.e., deciding whether a given property $\psi$ holds in all the reachable states, and the dual of safety, i.e., *reachability*, deciding if there exists a trajectory starting from the initial set that reaches a state satisfying the given property $\psi$. In principle, safety verification or reachability analysis aims to show that all trajectories of the hybrid systems starting from the initial set cannot enter some unsafe regions in the state space.

Safety verification or reachability analysis of hybrid systems presents a more difficult challenge, primarily due to the infinite number of possible states in continuous state space. Some well-established techniques have been proposed. In [2, 11], quantifier elimination was used to calculate exact reachable sets for linear systems with certain eigenstructures and semialgebraic initial sets. Tiwari [28] generalized this method to handle linear systems with almost arbitrary eigenstructures. In [5, 30, 10], level set methods, ellipsoidal techniques and flow-pipe approximations have been presented for computing approximate reachable sets of hybrid systems.

Recently, some methods [20, 21, 25, 28] based on invariant generation have been proposed for safety verification of hybrid systems. An invariant [24] of a hybrid system is a property that holds in all the reachable states of the system, in other words, it is an over-approximation of all the reachable states of the system. Invariants are useful facts about the dynamics of a given system, and are widely used in numerous approaches to analyze and verify systems. For example, if the invariants lie inside the safe regions, or their intersection with the unsafe regions is empty, then safety of hybrid systems is verified.

The problem of generating invariants of an arbitrary form is known to be computationally hard, intractable even for the simplest classes. The usual technique for generating invariants is to produce an *inductive invariant*, i.e., an assertion that holds at the initial states of the system, and is preserved by all discrete and continuous state changes. There has been a considerable volume of work towards invariant generation for hybrid systems using techniques in convex optimization, semi-algebraic system solving [4, 7, 13, 15, 17, 18, 20, 24, 25, 27, 29]. However, some of these techniques are only applicable to linear systems, some are subject to numerical errors and some suffer from high complexity. In virtue of the efficiency of numerical computation and the error-free property of symbolic computation, a hybrid symbolic-numeric method via SOS relaxation and exact certificate is presented in [31] to construct inequality invariants for continuous dynamic systems given by nonlinear vector fields.

In this work, we study how to generate inequality invariants for safety verification of nonlinear hybrid systems. We present a hybrid symbolic-numeric method, based on sum-of-squares (SOS) relaxation via semidefinite programming (SDP) and exact SOS representation recovery, to generate inequality invariants of hybrid systems, which guarantee that all the reachable states never enter the given unsafe regions. The idea is as follows: (1) Given a safe property, we predeterminate the templates of the invariants, and construct a semidefinite programming (SDP) system to solve the corresponding parametric polynomial optimization problem. (2) An exact invariant is obtained by recovering the exact SOS representation from the approximate solution of the associated SDP system. In the recovery step, Gauss-Newton iteration is deployed to refine the approximate solution from SDP solver. Then safety property of the hybrid systems can be easily verified, by the exact SOS representations of the conditions of the invariants. More details will be shown in Section 3.

Unlike the numerical approaches, our method can yield exact invariants, which can overcome the unsoundness in the verification of hybrid systems caused by numerical errors [19]. In comparison with some symbolic approaches of invariant generation based on qualifier elimination technique, our approach is more efficient and practical, because parametric polynomial optimization problem, based on SOS relaxation method, can be solved in polynomial time theoretically.

The rest of the paper is organized as follows. In Section 2, we introduce some notions about hybrid systems and invariants. Section 3 is devoted to illustrating a symbolic-numeric approach to generate invariants for safety verification of hybrid systems. In Section 4, we present some examples on invariant generation for safety verification of hybrid systems. Section 5 concludes the paper and discusses some future work.

## 2. Invariants

To model hybrid systems, we recall the definition of hybrid automata [8, 25].

**Definition 1** (Hybrid system). *A* hybrid system $\mathbf{H} : \langle V, L, \mathcal{T}, \Theta, \mathcal{D}, \Psi, \ell_0 \rangle$ *consists of the following components:*

- $V = \{x_1, ..., x_n\}$, *a set of real-valued system* variables. *A* state *is an interpretation of* $V$, *assigning to each* $x_i \in V$ *a real value. An* assertion *is a first-order formula over* $V$. *A state* $s$ *satisfies an assertion* $\varphi$, *written as* $s \models \varphi$, *if* $\varphi$ *holds on the state* $s$. *We will also write* $\varphi_1 \models \varphi_2$ *for two assertions* $\varphi_1, \varphi_2$ *to denote that* $\varphi_2$ *is true at least in all the states in which* $\varphi_1$ *is true;*

- $L$, *a finite set of locations;*

- $\mathcal{T}$, *a set of (discrete) transitions. Each transition* $\tau : \langle \ell, \ell', g_\tau, \rho_\tau \rangle \in \mathcal{T}$ *consists of a prelocation* $\ell \in L$, *a postlocation* $\ell' \in L$, *the guard condition* $g_\tau$ *over* $V$, *and an assertion* $\rho_\tau$ *over* $V \cup V'$ *representing the next-state relation, where* $V' = \{x'_1, ..., x'_n\}$ *denotes the next-state variables. Note that the transition* $\tau$ *can take place only if* $g_\tau$ *holds;*

- $\Theta$, *an assertion specifying the* initial *condition;*

- $\mathcal{D}$, *a map that maps each location* $\ell \in L$ *to a* differential rule *(also known as a* vector field *or a* flow field*)* $\mathcal{D}(\ell)$, *of the form* $\dot{x}_i = f_{\ell,i}(V)$ *for each* $x_i \in V$, *written briefly as* $\dot{\mathbf{x}} = \mathbf{f}_\ell(\mathbf{x})$. *The differential rule at a location specifies how the system variables evolve in that location;*

- $\Psi$, *a map that maps each location* $\ell \in L$ *to a* location condition (location invariant) $\Psi(\ell)$, *an assertion over* $V$*;*

- $\ell_0 \in L$, *the* initial condition. *We assume that the initial condition satisfies the location invariant at the initial location, that is,* $\Theta \models \Psi(\ell_0)$.

By a *state* of a hybrid system $\mathbf{H} : \langle V, L, \mathcal{T}, \Theta, \mathcal{D}, \Psi, \ell_0 \rangle$, we mean the tuple $(\ell, \mathbf{x}) \in L \times \mathbb{R}^n$ where $n$ is the number of program variables in $\mathbf{H}$.

**Definition 2** (Computation)**.** *[25] A* computation *of a hybrid system* $\mathbf{H}$ *is an infinite sequence of states*

$$< l_0, \mathbf{x}_0 >,\ < l_1, \mathbf{x}_1 >, \cdots, < l_i, \mathbf{x}_i >,\ < l_{i+1}, \mathbf{x}_{i+1} >, \cdots$$

*such that*

- **[Initiation]** $l_0 = \ell_0$ *and* $\mathbf{x}_0 \models \Theta$*;*

  *Furthermore, for each consecutive pair* $< l_i, \mathbf{x}_i >, < l_{i+1}, \mathbf{x}_{i+1} >$*, one of the two* consecution *conditions holds:*

- **[Discrete Consecution]** *There exists a transition* $\tau : \langle \ell, \ell', g_\tau, \rho_\tau \rangle$ *such that* $l_i = \ell, l_{i+1} = \ell'$ *and* $(\mathbf{x}_i, \mathbf{x}_{i+1}) \models \rho_\tau(\mathbf{x}_i, \mathbf{x}_{i+1})$ *if* $g_\tau$ *holds, or*

- **[Continuous Consecution]** $l_i = l_{i+1} = \ell$*, and there exists a time interval* $\delta > 0$ *and a smooth (continuous and differentiable to all orders) function* $f : [0, \delta] \to \mathbb{R}^n$ *s.t.* $f$ *evolves from* $\mathbf{x}_i$ *to* $\mathbf{x}_{i+1}$ *according to the differential rule* $\mathcal{D}(\ell)$ *at location* $\ell$*, while satisfying the location invariant* $\Psi(\ell)$*. Formally,*

  - $f(0) = \mathbf{x}_i, f(\delta) = \mathbf{x}_{i+1}$ *and* $\forall t \in [0, \delta], f(t) \models \Psi(\ell)$,
  - $\forall t \in [0, \delta), (f(t), \dot{f}(t)) \models \mathcal{D}(\ell)$.

A state $\langle \ell, \mathbf{x} \rangle$ is a *reachable* state of a hybrid system $\mathbf{H}$ if it appears in a computation of $\mathbf{H}$.

Figure 1 is a graphical representation of a hybrid system with two locations $\ell_1, \ell_2$. A state of this hybrid system is denoted by $\langle \ell, \mathbf{x} \rangle \in \{\ell_1, \ell_2\} \times \mathbb{R}^n$, and the initial state set is $\ell_1 \times \Theta$. During a continuous flow, the discrete location $\ell_i$ is maintained and the continuous state variables $\mathbf{x}$ evolve according to the differential equations $\dot{\mathbf{x}} = f_{\ell_i}(\mathbf{x})$, with $\mathbf{x}$ satisfying the location invariant $\Psi(\ell_i)$. At the state $\langle \ell_i, \mathbf{x} \rangle$, if the guard condition $g(\ell_i, \ell_j)$ is met, the system may undergo a transition to location $\ell_j$, and $\mathbf{x}$ will take the new value $\mathbf{x}'$, which is determined by the reset map $\rho(\ell_i, \ell_j)$.

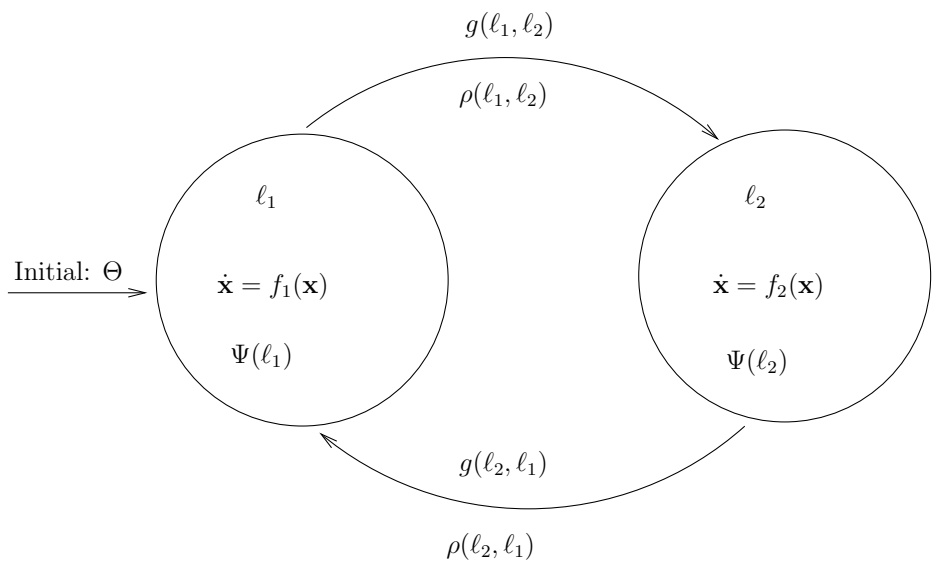

Figure 1: An example of hybrid system **H**

Given a hybrid system with an initial set and a prespecified safe (or unsafe) region, the system is *safe* if starting from any state in the initial set, this system would never evolve to the given unsafe region or the system would always stay inside the safe region. More specifically, consider the hybrid system **H** shown in Figure 1 and let $X_u \subset \mathbb{R}^n$ be an unsafe region. The system **H** is said to be safe if all trajectories of the system starting from any state in $\langle \ell_1, \mathbf{x}_0 \rangle \in \ell_1 \times \Theta$ can not reach $X_u$, or any state in $X_u$ is not reachable.

In this work, we will apply the invariant generation method to verify safety of hybrid systems. The following definitions of invariants of hybrid systems come from [25].

**Definition 3** (Invariant)**.** *An* invariant *of a hybrid system at location $\ell$ is an assertion $\mathcal{I}$ such that for any reachable state $\langle \ell, \mathbf{x} \rangle$ of the hybrid system, $\mathbf{x} \models \mathcal{I}$.*

*An* invariant *of a hybrid system is an assertion that holds in all the reachable states of the system.*

The problem to generate invariants with arbitrary form is known to be computationally hard, intractable even for the simplest classes. The usual technique for generating invariants is to compute inductive invariants, defined as follows.

**Definition 4** (Inductive invariant)**.** *An* inductive assertion map $\mathcal{I}$ *of a hybrid system* $\mathbf{H} : \langle V, L, \mathcal{T}, \Theta, \mathcal{D}, \Psi, \ell_0 \rangle$ *is a map that associates with each location $\ell \in L$ an assertion $\mathcal{I}(\ell)$ that holds initially and is preserved by all* *discrete transitions and continuous flows of* $\mathbf{H}$*. More formally, an inductive assertion map satisfies the following requirements:*

**(i)** [**Initial**] $\Theta \models \mathcal{I}(\ell_0)$.

**(ii)** [**Discrete Consecution**] *For each discrete transition* $\tau : \langle \ell, \ell', g_\tau, \rho_\tau \rangle$*, starting from a state satisfying* $\mathcal{I}(\ell)$*, and taking* $\tau$ *leads to a state satisfying* $\mathcal{I}(\ell')$*. Formally,*

$$\mathcal{I}(\ell) \wedge g_\tau \wedge \rho_\tau \models \mathcal{I}(\ell')$$

*where* $\mathcal{I}(\ell')$ *represents the assertion* $\mathcal{I}(\ell)$ *with the current state variables* $x_1, \ldots, x_n$ *replaced by the next state variables* $x'_1, \ldots, x'_n$*, respectively.*

**(iii)** [**Continuous Consecution**] *For every location* $\ell \in L$ *and states* $\langle \ell, \mathbf{x}_1 \rangle$*,* $\langle \ell, \mathbf{x}_2 \rangle$ *such that* $\mathbf{x}_2$ *evolves from* $\mathbf{x}_1$ *according to the differential rule* $\mathcal{D}(\ell)$ *at* $\ell$*, if* $\mathbf{x}_1 \models \mathcal{I}(\ell)$ *then* $\mathbf{x}_2 \models \mathcal{I}(\ell)$*.*

Our definition of inductive invariants is slightly modified from that of Definition 4 in [25], and the only change made is taking the guard conditions into account.

For a hybrid system, a formula $\mathcal{I}(\mathbf{x})$ is called a *differential invariant* at location $\ell$ if $\mathcal{I}(\mathbf{x})$ satisfies conditions (i) and (iii), that is, $\mathcal{I}(\mathbf{x})$ holds initially and is preserved by the continuous flow at a single location. There are several literature to compute differential invariants. [13] presented an approach based on the computable algebraic-geometry theory to generate differential invariants. [20] computed differential invariants using a verification logic for hybrid systems. [31] suggested a hybrid symbolic-numeric method to compute inequality differential invariants.

**Remark 1.** *Clearly, inductive invariants are over-approximation of the reachable sets of hybrid systems, since an inductive invariant is true for all the reachable states of the system.*

## 3. Safety Verification of Hybrid Systems

The aim of this section is to translate the problem of safety verification of hybrid systems into that of generating invariants, which can be transformed further into polynomial optimization problem with parameters. We will present a hybrid symbolic-numeric method, based on SOS relaxation, to solve this polynomial optimization problem, and obtain the invariants, which can guarantee the safety property of hybrid systems.

*3.1. Invariants and Safety Verification*

In this paper, we are interested in hybrid systems in which the relations are given by (real) polynomials over the system variables. Then we define

**Definition 5** (Polynomial Hybrid System)**.** *A polynomial hybrid system is a hybrid system:* $\mathbf{H} : \langle V, L, \mathcal{T}, \Theta, \mathcal{D}, \Psi, \ell_0 \rangle$*, where*

- *for each transition* $\tau : \langle \ell, \ell', g_\tau, \rho_\tau \rangle \in \mathcal{T}$*, the guard condition* $g_\tau$ *(resp. the reset relation* $\rho_\tau$*) is a conjunction of polynomial inequalities over* $V$ *(resp.* $V \cup V'$*); also, the initial condition* $\Theta$ *and the location invariant* $\Psi(\ell)$*, for each* $\ell \in L$*, are conjunctions of polynomial inequalities over* $V$*;*
- *each rule* $\mathcal{D}(\ell)$ *is of the form* $\dot{x}_i = f_{\ell,i}(\mathbf{x})$ *for each* $x_i \in V$*, where* $f_{\ell,i}(\mathbf{x}) \in \mathbb{R}[\mathbf{x}]$*.*

We are interested in finding invariants of the form $\varphi_\ell(\mathbf{x}) \geq 0$ at location $\ell \in L$. Below is an alternative expression of Definition 4.

**Theorem 1.** *Let* $\mathbf{H} : \langle V, L, \mathcal{T}, \Theta, \mathcal{D}, \Psi, \ell_0 \rangle$ *be a hybrid system. Suppose for each location* $\ell \in L$*, there exists a function* $\varphi_\ell(\mathbf{x})$ *satisfying the following conditions:*

**(i)** $\Theta \models \varphi_{\ell_0}(\mathbf{x}) \geq 0$,

**(ii)** $\varphi_\ell(\mathbf{x}) \geq 0 \wedge g(\ell, \ell') \wedge \rho(\ell, \ell') \models \varphi_{\ell'}(\mathbf{x}') \geq 0$*, for any transition* $\langle \ell, \ell', g, \rho \rangle$ *going out from* $\ell$*,*

**(iii)** $\varphi_\ell(\mathbf{x}) \geq 0 \wedge \Psi(\ell) \models \dot{\varphi}_\ell(\mathbf{x}) > 0$*, here* $\dot{\varphi}_\ell(\mathbf{x})$ *denotes the* Lie-derivative *of* $\varphi_\ell$ *along the vector field* $\mathcal{D}(\ell)$*, i.e.,* $\dot{\varphi}_\ell(\mathbf{x}) = \sum_{i=1}^{n} \frac{\partial \varphi_\ell}{\partial x_i} \cdot f_{\ell,i}(\mathbf{x})$*,*

*then* $\varphi_\ell(\mathbf{x}) \geq 0$ *is an invariant of the hybrid system* $\mathbf{H}$ *at location* $\ell$*.*

*Proof.* The proof follows directly from Definition 4. □

Remarked that if the functions $\varphi_\ell(\mathbf{x})$ at all locations are identical to $\varphi(\mathbf{x})$, then $\varphi(\mathbf{x})$ is an inductive invariant of the given hybrid system, as described in the following theorem.

**Theorem 2.** *Let* $\mathbf{H}$ *be a hybrid system. Suppose there exists a function* $\varphi(\mathbf{x})$ *satisfying the following conditions:*

**(i)** $\Theta \models \varphi(\mathbf{x}) \geq 0$,

**(ii)** $\varphi(\mathbf{x}) \geq 0 \wedge g(\ell, \ell') \wedge \rho(\ell, \ell') \models \varphi(\mathbf{x}') \geq 0$, *for any transition* $\langle \ell, \ell', g, \rho \rangle$ *going out from* $\ell$,

**(iii)** $\varphi(\mathbf{x}) \geq 0 \wedge \Psi(\ell) \models \dot{\varphi}(\mathbf{x}) > 0$,

*then* $\varphi(\mathbf{x}) \geq 0$ *is an inductive invariant of the system* $\mathbf{H}$.

In the sequel, for brevity, we shall use $\varphi_\ell(\mathbf{x})$ to denote both the invariant $\varphi_\ell(\mathbf{x}) \geq 0$ and $\varphi_\ell(\mathbf{x})$.

The following theorem shows that invariants can be applied to verify the safety property of hybrid systems.

**Theorem 3.** *Let* $\mathbf{H}$ *be a hybrid system, and* $X_u(\ell)$ *be the unsafe region at location* $\ell$. *Suppose there exists functions* $\varphi_\ell(\mathbf{x})$, *for* $\ell \in L$, *that satisfy the conditions (i-iii) in Theorem 1, and moreover,*

**(iv)** $X_u(\ell) \models \varphi_\ell(\mathbf{x}) < 0, \quad \forall \ell \in L$,

*then the safety of the system* $\mathbf{H}$ *is guaranteed.*

*Proof.* Clearly, $\varphi_\ell(\mathbf{x}) \geq 0$ is an invariant of hybrid system $\mathbf{H}$ at location $\ell$. Then the condition (iv) implies that all reachable sets at location $\ell$ lie outside the unsafe region $X_u(\ell)$, yielding the safety of the system. □

Similarly, inductive invariants can be applied to verify safety of hybrid systems.

**Theorem 4.** *Let* $\mathbf{H}$ *be a hybrid system, and* $X_u(\ell)$ *be the unsafe region at location* $\ell$. *Suppose there exists a function* $\varphi(\mathbf{x})$ *that satisfies the conditions (i-iii) in Theorem 2, and moreover,*

**(iv)** $X_u(\ell) \models \varphi(\mathbf{x}) < 0, \quad \forall \ell \in L$,

*then the safety of the system is guaranteed.*

*Proof.* The proof is similar to that of Theorem 3. □

**Remark 2.** *Functions* $\varphi_\ell(\mathbf{x})$ *and* $\varphi(\mathbf{x})$ *in Theorems 3 and 4 are also known as barrier certificates in [21].*

*3.2. Sum of Squares Relaxation*

According to Theorems 3 and 4, to verify the safety of hybrid system **H**, it suffices to compute real polynomials $\varphi_\ell(\mathbf{x})$ or $\varphi(\mathbf{x})$.

In the following, we only discuss how to find the invariant $\varphi_\ell(\mathbf{x})$ at each location $\ell \in L$. The problem of computing the inductive invariant $\varphi(\mathbf{x})$ can be handled similarly.

Our idea of computing $\varphi_\ell(\mathbf{x})$ or $\varphi(\mathbf{x})$, based on Sum-of-Squares (SOS) relaxation and rational vector recovery, is as follows.

**Step 1:** Predetermine a template of polynomial invariants with the given degree and convert the problem of computing polynomial invariants to the associated parametric polynomial optimization problem. SOS relaxation method is then applied to obtain a polynomial invariant with floating point coefficients.

**Step 2:** Apply Gauss-Newton refinement and rational vector recovery on the approximate polynomial invariant to get polynomials with rational coefficients, which exactly satisfy the conditions of invariants of the given hybrid system.

The problem of computing the invariant $\varphi_\ell(\mathbf{x})$ at each location $\ell \in L$, that satisfy the conditions in Theorem 3 can be transformed into the following problem

$$\left.\begin{array}{l} \text{find } \varphi_\ell(\mathbf{x}) \in \mathbb{R}[\mathbf{x}], \forall \ell \in L \\ \quad \text{s.t.} \quad \Theta \models \varphi_{\ell_0}(\mathbf{x}) \geq 0, \\ \qquad\quad \varphi_\ell(\mathbf{x}) \geq 0 \wedge g(\ell, \ell') \wedge \rho(\ell, \ell') \models \varphi_{\ell'}(\mathbf{x}') \geq 0, \\ \qquad\quad \varphi_\ell(\mathbf{x}) \geq 0 \wedge \Psi(\ell) \models \dot{\varphi}_\ell(\mathbf{x}) > 0, \\ \qquad\quad X_u(\ell) \models \varphi_\ell(\mathbf{x}) < 0. \end{array}\right\} \tag{1}$$

Let us first predetermine a template of polynomial invariants with the given degree $d$, that is, we assume

$$\varphi_\ell(\mathbf{x}) = \sum_\alpha \mathbf{c}_\alpha \mathbf{x}^\alpha, \tag{2}$$

where $\mathbf{x}^\alpha = x_1^{\alpha_1} \cdots x_n^{\alpha_n}$ and $\mathbf{c}_\alpha \in \mathbb{R}$ are parameters, with $\alpha \in \mathbb{Z}_{\geq 0}^n$ and $\sum_{i=1}^n \alpha_i \leq d$. One can apply quantifier elimination methods to solve the corresponding parametric semi-algebraic systems, and for the given template, quantifier elimination methods can yield the sufficient and necessary conditions for the existence of invariants. Several Maple packages, such as RAGLib[16] and DISCOVERER [32], are available to solve this problem. However, quantifier elimination method based on the cylindrical algebraic decomposition (CAD) are of high complexity. Instead, we will explore the SOS relaxation techniques based on semidefinite programming (SDP) solving to obtain polynomial invariants.

In the sequel, we suppose that

$$\Theta = \{\mathbf{x} \in \mathbb{R}^n : \bigwedge_{l=1}^{q} \theta_l(\mathbf{x}) \geq 0\},\ X_u(\ell) = \{\mathbf{x} \in \mathbb{R}^n : \bigwedge_{j=1}^{p} \zeta_{\ell,j}(\mathbf{x}) \geq 0\},$$

$$\Psi(\ell) = \{\mathbf{x} \in \mathbb{R}^n : \bigwedge_{k=1}^{r} \psi_{\ell,k}(\mathbf{x}) \geq 0\},\ g(\ell,\ell') = \{\mathbf{x} \in \mathbb{R}^n : \bigwedge_{i=1}^{s} g_{\ell\ell',i}(\mathbf{x}) \geq 0\},$$

$$\rho(\ell,\ell')(\mathbf{x},\mathbf{x}') = \{\mathbf{x}' \in \mathbb{R}^n : \bigwedge_{u=1}^{t} \rho_{\ell\ell',u}(\mathbf{x},\mathbf{x}') \geq 0\},$$

where $\ell, \ell' \in L$, and $\theta_l(\mathbf{x})$, $\zeta_{\ell,j}(\mathbf{x})$, $\psi_{\ell,k}(\mathbf{x})$, $g_{\ell\ell',i}(\mathbf{x})$ and $\rho_{\ell\ell',u}(\mathbf{x},\mathbf{x}')$ are polynomials.

Clearly, a sufficient condition for $r(\mathbf{x}) \in \mathbb{R}[\mathbf{x}]$ with degree $2e$ to be positive semidefinite is that there exists an SOS of $r(\mathbf{x})$:

$$r(\mathbf{x}) = \sum_i {r_i}^2(\mathbf{x}), \quad \text{with } r_i(\mathbf{x}) \in \mathbb{R}[\mathbf{x}], \tag{3}$$

or, equivalently, $r(\mathbf{x})$ can be represented as

$$r(\mathbf{x}) = \mathbf{m}(\mathbf{x})^T \cdot W \cdot \mathbf{m}(\mathbf{x}),$$

where $W$ is a real symmetric and positive semidefinite matrix, and $m(\mathbf{x})$ is a vector of terms in $\mathbb{R}[\mathbf{x}]$ with degree $\leq e$.

When a polynomial $r(\mathbf{x})$ can be written as an SOS in $\mathbb{R}[\mathbf{x}]$, we simply call $r(\mathbf{x})$ an SOS. Denote by $\Sigma_{n,2e}$ the set of all SOSes of degree $\leq 2e$ in variables $x_1, ..., x_n$, i.e.,

$$\Sigma_{n,2e} = \{r(\mathbf{x}) \in \mathbb{R}[\mathbf{x}] : r(\mathbf{x}) \,\text{is an SOS},\, \deg(r(\mathbf{x})) \leq 2e\}.$$

Based on the SOS relaxation, the constraints in (1) can be replaced by stronger ones. For instance, to find a polynomial $\varphi_{\ell_0}(\mathbf{x})$ satisfying

$$\Theta \models \varphi_{\ell_0}(\mathbf{x}) \geq 0$$

it suffices to find $\varphi_{\ell_0}(\mathbf{x})$ such that

$$\varphi_{\ell_0}(\mathbf{x}) = \sigma_0(\mathbf{x}) + \sum_{l=1}^{q} \sigma_l(\mathbf{x})\theta_l(\mathbf{x}),$$

where $\sigma_0, \sigma_l \in \mathbb{R}[\mathbf{x}]$ are SOSes. Therefore, the problem of computing polynomials $\varphi_\ell(\mathbf{x})$ is transformed into the following SOS program:

$$\left.\begin{array}{ll}
\text{find} & \varphi_\ell(\mathbf{x}) \in \mathbb{R}[\mathbf{x}], \forall \ell \in L \\
\text{s.t.} & \varphi_{\ell_0}(\mathbf{x}) = \sigma_0(\mathbf{x}) + \sum_{l=1}^{q} \sigma_l(\mathbf{x})\theta_l(\mathbf{x}), \\
& \varphi_{\ell'}(\mathbf{x}') = \lambda_{\ell\ell',0}(\mathbf{x}) + \sum_{i=1}^{s} \lambda_{\ell\ell',i}(\mathbf{x}) g_{\ell\ell',i}(\mathbf{x}) \\
& \qquad + \sum_{u=1}^{t} \gamma_{\ell\ell',u}(\mathbf{x})\rho_{\ell\ell',u}(\mathbf{x},\mathbf{x}') + \eta_{\ell\ell'}(\mathbf{x})\varphi_\ell(\mathbf{x}), \\
& \dot{\varphi}_\ell(\mathbf{x}) = \phi_{\ell,0}(\mathbf{x}) + \sum_{k=1}^{r} \phi_{\ell,k}(\mathbf{x})\psi_{\ell,k}(\mathbf{x}) + \nu_\ell(\mathbf{x})\varphi_\ell(\mathbf{x}) + \epsilon_{\ell,1} \\
& -\varphi_\ell(\mathbf{x}) = \mu_{\ell,0}(\mathbf{x}) + \sum_{j=1}^{p} \mu_{\ell,j}(\mathbf{x})\zeta_{\ell,j}(\mathbf{x}) + \epsilon_{\ell,2},
\end{array}\right\} \quad (4)$$

where $\sigma_l(\mathbf{x}), \lambda_{\ell\ell',i}(\mathbf{x}), \gamma_{\ell\ell',u}(\mathbf{x}), \eta_{\ell\ell'}(\mathbf{x}), \phi_{\ell,k}(\mathbf{x}), \nu_\ell(\mathbf{x}), \mu_{\ell,j}(\mathbf{x}) \in \Sigma_{n,2e}$ and $\epsilon_{\ell,1}$, $\epsilon_{\ell,2} \in \mathbb{R}_+$. The decision variables are the coefficients of all polynomials appearing in (4), such as $\varphi_\ell(\mathbf{x}), \sigma_l(\mathbf{x}), \lambda_{\ell\ell',i}(\mathbf{x})$.

Since the coefficients of $\varphi_\ell(\mathbf{x}), \eta_{\ell\ell'}(\mathbf{x})$ and $\nu_\ell(\mathbf{x})$ are unknown, some nonlinear terms that are products of these coefficients, occur in the second and third constraints of (4). The SOS relaxation will then lead to a non-convex bilinear matrix inequalities (BMI) problem. To avoid BMI problem, we adopt stronger conditions to compute the invariants of hybrid systems.

**Theorem 5.** *Under the assumptions in Theorem 1, suppose for each $\ell \in L$, $\varphi_\ell(\mathbf{x})$ satisfies the following conditions:*

**(i)** $\Theta \models \varphi_{\ell_0}(\mathbf{x}) \geq 0$,

**(ii')** $g(\ell,\ell') \wedge \rho(\ell,\ell') \models \varphi_{\ell'}(\mathbf{x}') \geq 0$, *for any transition $\langle \ell, \ell', g, \rho \rangle$ going out from $\ell$,*

**(iii')** $\Psi(\ell) \models \dot{\varphi}_\ell(\mathbf{x}) > 0$,

*then $\varphi_\ell(\mathbf{x}) \geq 0$ is an invariant of the hybrid system $\mathbf{H}$ at location $\ell$. In addition, if $\varphi_\ell(\mathbf{x})$ satisfies*

**(iv)** $X_u(\ell) \models \varphi_\ell(\mathbf{x}) < 0, \forall \ell \in L$,

*then the safety of the system is guaranteed.*

*Proof.* Since the conditions **(ii')** and **(iii')** are stronger than the conditions **(ii)** and **(iii)** in Theorem 1 respectively, $\varphi_\ell$ is an invariant at location $\ell$. According to Theorem 3, the condition **(iv)** can guarantee the safety of this system. □

A similar conclusion can be attained for inductive invariants, as stated in the following

**Theorem 6.** *Under the assumptions in Theorem 2, suppose there exists a polynomial $\varphi(\mathbf{x})$ satisfying the following conditions:*

**(i)** $\Theta \models \varphi(\mathbf{x}) \geq 0$,

**(ii')** $g(\ell, \ell') \wedge \rho(\ell, \ell') \models \varphi(\mathbf{x}') \geq 0$, *for any transition* $\langle \ell, \ell', g, \rho \rangle$ *going out from* $\ell$,

**(iii')** $\Psi(\ell) \models \dot{\varphi}(\mathbf{x}) > 0, \forall \ell \in L$,

*then $\varphi(\mathbf{x}) \geq 0$ is an inductive invariant of the hybrid system $\mathbf{H}$. In addition, if $\varphi(\mathbf{x})$ satisfies*

**(iv)** $X_u(\ell) \models \varphi(\mathbf{x}) < 0, \forall \ell \in L$,

*then the safety of the system is guaranteed.*

Having Theorem 5, the program (4) can be modified into the following problem:

$$\left.\begin{array}{ll}
\text{find} & \varphi_\ell(\mathbf{x}) \in \mathbb{R}[\mathbf{x}], \forall \ell \in L \\
\text{s.t.} & \varphi_{\ell_0}(\mathbf{x}) = \sigma_0(\mathbf{x}) + \sum_{l=1}^{q} \sigma_l(\mathbf{x})\theta_l(\mathbf{x}), \\
& \varphi_{\ell'}(\mathbf{x}') = \lambda_{\ell\ell',0}(\mathbf{x}) + \sum_{i=1}^{s} \lambda_{\ell\ell',i}(\mathbf{x}) g_{\ell\ell',i}(\mathbf{x}) + \sum_{u=1}^{t} \gamma_{\ell\ell',u}(\mathbf{x})\rho_{\ell\ell',u}(\mathbf{x}, \mathbf{x}'), \\
& \dot{\varphi}_\ell(\mathbf{x}) = \phi_{\ell,0}(\mathbf{x}) + \sum_{k=1}^{r} \phi_{\ell,k}(\mathbf{x})\psi_{\ell,k}(\mathbf{x}) + \epsilon_{\ell,1}, \\
& -\varphi_\ell(\mathbf{x}) = \mu_{\ell,0}(\mathbf{x}) + \sum_{j=1}^{p} \mu_{\ell,j}(\mathbf{x})\zeta_{\ell,j}(\mathbf{x}) + \epsilon_{\ell,2},
\end{array}\right\} \tag{5}$$

where $\sigma_l(\mathbf{x}), \lambda_{\ell\ell',i}(\mathbf{x}), \gamma_{\ell\ell',u}(\mathbf{x}), \phi_{\ell,k}(\mathbf{x}), \mu_{\ell,j}(\mathbf{x}) \in \Sigma_{n,2e}$ and $\epsilon_{\ell,1}$, $\epsilon_{\ell,2} \in \mathbb{R}_+$. The

program is equivalent to the following SDP problem:

$$
\left.\begin{array}{l}
\inf \quad \mathrm{Tr}(M,W,V,P,Q) \\
\text{s.t. } \varphi_{\ell_0}(\mathbf{x}) = \mathbf{m}_0(\mathbf{x})^T \cdot M^{[0]} \cdot \mathbf{m}_0(\mathbf{x}) + \sum_{l=1}^{q} \mathbf{m}_l(\mathbf{x})^T \cdot M^{[l]} \cdot \mathbf{m}_l(\mathbf{x})\theta_l(\mathbf{x}), \\
\quad \varphi_{\ell'}(\mathbf{x}') = \mathbf{w}_{\ell\ell',0}(\mathbf{x})^T \cdot W^{[\ell\ell',0]} \cdot \mathbf{w}_{\ell\ell',0}(\mathbf{x}) \\
\qquad\qquad + \sum_{i=1}^{s} \mathbf{w}_{\ell\ell',i}(\mathbf{x})^T \cdot W^{[\ell\ell',i]} \cdot \mathbf{w}_{\ell\ell',i}(\mathbf{x}) g_{\ell\ell',i}(\mathbf{x}) \\
\qquad\qquad + \sum_{u=1}^{t} \mathbf{v}_{\ell\ell',u}(\mathbf{x})^T \cdot V^{[\ell\ell',u]} \cdot \mathbf{v}_{\ell\ell',u}(\mathbf{x}) \rho_{\ell\ell',u}(\mathbf{x},\mathbf{x}'), \\
\quad \dot{\varphi}_{\ell}(\mathbf{x}) = \mathbf{p}_{\ell,0}(\mathbf{x})^T \cdot P^{[\ell,0]} \cdot \mathbf{p}_{\ell,0}(\mathbf{x}) \\
\qquad\qquad + \sum_{k=1}^{r} \mathbf{p}_{\ell,k}(\mathbf{x})^T \cdot P^{[\ell,k]} \cdot \mathbf{p}_{\ell,k}(\mathbf{x}) \psi_{\ell,k}(\mathbf{x}) + \epsilon_{\ell,1}, \\
\quad \varphi_{\ell}(\mathbf{x}) = -\mathbf{q}_{\ell,0}(\mathbf{x})^T \cdot Q^{[\ell,0]} \cdot \mathbf{q}_{\ell,0}(\mathbf{x}) \\
\qquad\qquad - \sum_{j=1}^{p} \mathbf{q}_{\ell,j}(\mathbf{x})^T \cdot Q^{[\ell,j]} \cdot \mathbf{q}_{\ell,j}(\mathbf{x}) \zeta_{\ell,j}(\mathbf{x}) - \epsilon_{\ell,2},
\end{array}\right\} \tag{6}
$$

where all the matrices $M^{[l]}, W^{[\ell\ell',i]}, V^{[\ell\ell',u]}, P^{[\ell,k]}, Q^{[\ell,j]}$ are symmetric and positive semidefinite, and the function $\mathrm{Tr}(M,W,V,P,Q)$ denotes the sum of traces of all these matrices, which acts as a dummy objective function commonly used in SDP for optimization problem with no objective function.

Many Matlab packages of SDP solvers, such as SOSTOOLS [22], YALMIP [14], and SeDuMi [26], are available to solve the problem (6) efficiently.

### *3.3. Exact Certificate of Sum of Squares Decomposition*

Since the SDP solvers in Matlab is running in fixed precision, the techniques in Section 3.2 will yield numerical solutions to the associated SDP problem (6), where the numerical polynomial $\varphi_\ell(\mathbf{x})$ and numerical positive semidefinite matrices $M^{[l]}, \ldots, Q^{[\ell,j]}$ satisfy the constraints in (6) *approximately*, for instance,

$$
\varphi_{\ell_0}(\mathbf{x}) \approx \mathbf{m}_0(\mathbf{x})^T \cdot M^{[0]} \cdot \mathbf{m}_0(\mathbf{x}) + \sum_{l=1}^{q} \mathbf{m}_l(\mathbf{x})^T \cdot M^{[l]} \cdot \mathbf{m}_l(\mathbf{x})\theta_l(\mathbf{x}), \;\; M^{[l]} \gtrapprox 0. \tag{7}
$$

However, due to round-off errors, $\varphi_\ell(\mathbf{x}) \geq 0$ may not necessarily be an invariant of the given hybrid system at location $\ell$, because the constraints in (6) may not hold exactly, for example, (7) means that $\varphi_{\ell_0}(\mathbf{x})$ may not be positive semidefinite exactly within the initial set $\Theta$. Therefore in the next step, from the numerical polynomials $\varphi_\ell(\mathbf{x})$ and the numerical positive semidefinite matrices $M^{[l]}, \ldots, Q^{[\ell,j]}$, we will recover polynomials $\widetilde{\varphi}_\ell(\mathbf{x})$ with rational coefficients, which satisfy (6) *exactly*.

As described in [9], finding a polynomial with rational coefficients can be translated into the problem of rational vector recovery. In Section 3.2, a numerical vector $\mathbf{v}_\ell^*$ denoting the coefficient (column) vector of $\varphi_\ell(\mathbf{x})$ is obtained by solving the SDP system, i.e., $\varphi_\ell(\mathbf{x}) = \mathbf{v}_\ell^{*T} \cdot T_\ell(\mathbf{x})$, where $T_\ell(\mathbf{x})$ is the column vector of all terms in $\varphi_\ell(\mathbf{x})$. To obtain a rational vector $\tilde{\mathbf{v}}_\ell$ near to $\mathbf{v}_\ell^*$, we can employ the simultaneous Diophantine approximation algorithm [12], once the bound of the common denominator of $\tilde{\mathbf{v}}_\ell$ is given.

The recovery of the matrices $M^{[l]}, \ldots, Q^{[\ell,j]}$ into rational positive semidefinite matrices is split into two steps. We first recover the matrices $\widetilde{M}^{[l]}, \ldots, \widetilde{Q}^{[\ell,j]}$ for $1 \le l \le q, \ldots, 1 \le j \le p$ and then recover $\widetilde{M}^{[0]}, \ldots, \widetilde{Q}^{[\ell,0]}$. To illustrate the idea, we only discuss how to recover $M^{[l]}$ for $1 \le l \le q$ and the matrices $W^{[\ell\ell',i]}, V^{[\ell\ell',u]}, P^{[\ell,k]}, Q^{[\ell,j]}$ can be recovered similarly.

Given the numerical positive semidefinite matrices $M^{[l]}$, $1 \le l \le q$ in (6), we can find the nearby rational positive semidefinite matrices $\widetilde{M}^{[l]}$ by use of the rational vector recovery technique. In practice, all the $M^{[l]}$ are numerical diagonal matrices, in other words, the off-diagonal entries are very tiny and the diagonal entries are nonnegative. Therefore, by setting the small entries of $M^{[l]}$ to be zeros we easily get the nearby rational positive semidefinite matrices $\widetilde{M}^{[l]}$ for $l = 1, \ldots, q$. The nearby rational positive semidefinite matrices $\widetilde{W}^{[\ell\ell',i]}, \widetilde{V}^{[\ell\ell',u]}, \widetilde{P}^{[\ell,k]}, \widetilde{Q}^{[\ell,j]}$ can be recovered similarly.

Having $\widetilde{\varphi}_\ell(\mathbf{x}) = \tilde{\mathbf{v}}_\ell{}^T \cdot T_\ell(\mathbf{x})$ and $\widetilde{M}^{[l]}, \ldots, \widetilde{Q}^{[\ell,j]}$ for $1 \le l \le q, \ldots, 1 \le j \le p$, the program (6) is converted to

$$\left.\begin{array}{ll}\inf & \mathrm{Tr}(M^{[0]}, W^{[\ell\ell',0]}, P^{[\ell,0]}, Q^{[\ell,0]}) \\ & \text{s.t. } \widetilde{\varphi}_{\ell_0}(\mathbf{x}) - \sum_{l=1}^q \mathbf{m}_l(\mathbf{x})^T \cdot \widetilde{M}^{[l]} \cdot \mathbf{m}_l(\mathbf{x}) \theta_l(\mathbf{x}) \\ & \qquad \approx \mathbf{m}_0(\mathbf{x})^T \cdot M^{[0]} \cdot \mathbf{m}_0(\mathbf{x}), \\ & \widetilde{\varphi}_{\ell'}(\mathbf{x}') - \sum_{i=1}^s \mathbf{w}_{\ell\ell',i}(\mathbf{x})^T \cdot \widetilde{W}^{[\ell\ell',i]} \cdot \mathbf{w}_{\ell\ell',i}(\mathbf{x}) g_{\ell\ell',i}(\mathbf{x}) \\ & \qquad - \sum_{u=1}^t \mathbf{v}_{\ell\ell',u}(\mathbf{x})^T \cdot \widetilde{V}^{[\ell\ell',u]} \cdot \mathbf{v}_{\ell\ell',u}(\mathbf{x}) \rho_{\ell\ell',u}(\mathbf{x}, \mathbf{x}') \\ & \qquad \approx \mathbf{w}_{\ell\ell',0}(\mathbf{x})^T \cdot W^{[\ell\ell',0]} \cdot \mathbf{w}_{\ell\ell',0}(\mathbf{x}) \\ & \dot{\widetilde{\varphi}}_\ell(\mathbf{x}) - \sum_{k=1}^r \mathbf{p}_{\ell,k}(\mathbf{x})^T \cdot \widetilde{P}^{[\ell,k]} \cdot \mathbf{p}_{\ell,k}(\mathbf{x}) \psi_{\ell,k}(\mathbf{x}) - \widetilde{\epsilon}_{\ell,1} \\ & \qquad \approx \mathbf{p}_{\ell,0}(\mathbf{x})^T \cdot P^{[\ell,0]} \cdot \mathbf{p}_{\ell,0}(\mathbf{x}) \\ & \widetilde{\varphi}_\ell(\mathbf{x}) + \sum_{j=1}^p \mathbf{q}_{\ell,j}(\mathbf{x})^T \cdot \widetilde{Q}^{[\ell,j]} \cdot \mathbf{q}_{\ell,j}(\mathbf{x}) \zeta_{\ell,j}(\mathbf{x}) + \widetilde{\epsilon}_{\ell,2} \\ & \qquad \approx -\mathbf{q}_{\ell,0}(\mathbf{x})^T \cdot Q^{[\ell,0]} \cdot \mathbf{q}_{\ell,0}(\mathbf{x}) \end{array}\right\} \quad (8)$$

Observing (8), the matrices $M^{[0]}, \ldots, Q^{[\ell,0]}$ have floating point entries, while the matrices $\widetilde{M}^{[l]}, \ldots, \widetilde{Q}^{[\ell,j]}$ are rational positive semidefinite matrices. Therefore, the remaining task is to find nearby rational positive semidefinite matrices $\widetilde{M}^{[0]}, \ldots, \widetilde{Q}^{[\ell,0]}$ such that the constraints in (8) hold exactly. To fulfil this task, we can first apply Gauss-Newton iteration to refine $M^{[0]}, \ldots, Q^{[\ell,0]}$, and then recover the rational positive definite matrices $\widetilde{M}^{[0]}, \ldots, \widetilde{Q}^{[\ell,0]}$ respectively from the refined $M^{[0]}, \ldots, Q^{[\ell,0]}$, by orthogonal projection if the involved matrix is of full rank, or by rational vector recovery method otherwise.

Finally, we check if all the matrices $\widetilde{M}^{[0]}, \ldots, \widetilde{Q}^{[\ell,0]}$ are positive semidefinite. If so, then return $\widetilde{\varphi}_\ell(\mathbf{x}) \geq 0$ as an invariant of the given hybrid system at location $\ell \in L$; otherwise, return "we cannot find invariants of the given degree bound".

**Remark 3.** *The above technique based on SOS relaxation and exact polynomial recovery can be applied to computing the inductive invariants of hybrid systems, which guarantee the safety of the given hybrid system.*

*3.4. Algorithm*

The discussion in Section 3.3 leads to an algorithm of computing the (inductive) invariants of polynomial hybrid systems. As stated above, we only present how to compute the invariants $\varphi_\ell(\mathbf{x})$, for $\ell \in L$, that satisfy (6), and the case of computing the inductive invariants is similar.

*Algorithm Polynomial Inequality Invariant Generation*

Input:
- $\mathbf{H} : \langle V, L, \mathcal{T}, \Theta, \mathcal{D}, \Psi, \ell_0 \rangle$ a polynomial hybrid system.
- $d \in \mathbb{Z}_{>0}$: the degree bound of the candidate polynomial invariants.
- $D \in \mathbb{Z}_{>0}$: the bound of the common denominator of the coefficient vector of the polynomial invariants.
- $e \in \mathbb{Z}_{\geq 0}$: the degree bound $2e$ of the SOSes used to construct the SDP system.
- $\tau \in \mathbb{R}_{>0}$: the given tolerance.

Output:
- $\widetilde{\varphi}_\ell(\mathbf{x}) \geq 0$: the verified polynomia invariant at each location $\ell \in L$.

1. Compute the candidates of polynomial invariants
   (i) For each locaiton $\ell \in L$, predetermine the templates of $\varphi_\ell(\mathbf{x})$, with degree $d$, and construct an SDP system of the form (6), where the degree bounds of all the involved SOSes are $2e$.
   - If the SDP system (6) has no feasible solutions,
     return "we can't find polynomial invariants with degree $\leq d$ at each location";

- Otherwise,
  obtain a numerical vector $\mathbf{v}_\ell^*$, numerical constants $\epsilon_{\ell,1}$, $\epsilon_{\ell,2}$ and numerical positive semidefinite matrices $M^{[l]}$, $W^{[\ell\ell',i]}$, $V^{[\ell\ell',u]}$, $P^{[\ell,k]}$, $Q^{[\ell,j]}$ for $0 \le l \le q, 0 \le i \le s, 1 \le u \le t, 0 \le k \le r, 0 \le j \le p$.

(ii) For the common denominator bound $D$, compute from $\mathbf{v}_\ell^*$ a rational vector $\widetilde{\mathbf{v}}_\ell$ by Diophantine approximation algorithm, and get the associated rational polynomial $\widetilde{\varphi}_\ell(\mathbf{x})$. Similarly, the nearby positive contants $\widetilde{\epsilon}_{\ell,1}$ and $\widetilde{\epsilon}_{\ell,2}$ are obtained.

(iii) Convert all the $M^{[l]}, \ldots, Q^{[\ell,j]}$ into rational and positive semidefinite matrices $\widetilde{M}^{[l]}, \ldots, \widetilde{Q}^{[\ell,j]}$, for $1 \le l \le q, \ldots, 1 \le j \le p$.

2. Compute the exact SOS decomposition

(i) Reconstruct an SDP system of the form (8) to get approximate positive semidefinite matrices $M^{[0]}, \ldots, Q^{[\ell,0]}$ satisfying (8).

(ii) Apply Gauss-Newton iteration to refine the matrices $M^{[0]}, \ldots, Q^{[\ell,0]}$ obtained in Step 2 (i).

(iii) From the refined $M^{[0]}, \ldots, Q^{[\ell,0]}$, compute the rational matrices $\widetilde{M}^{[0]}, \ldots, \widetilde{Q}^{[\ell,0]}$ respectively by orthogonal projection method if the involved matrix is of full rank, or by rational vector recovery if the matrix is singular.

(iv) Check whether all the matrices $\widetilde{M}^{[0]}, \ldots, \widetilde{Q}^{[\ell,0]}$ are positive semidefinite.

- If so, return $\widetilde{\varphi}_\ell(\mathbf{x}) \ge 0$ as an invariant at location $\ell \in L$;
- Otherwise,
  return "we can't find polynomial invariants with degree $\le d$."

**Remark 4.** *Our algorithm cannot guarantee rational solutions will always be found since there exists limitations in the above algorithm on choosing the degree bound $e$ and the common denominator bound $D$. Furthermore, it is difficult to determine in advance whether there exists invariants with rational coefficients or not. Therefore, even if our algorithm cannot find the invariants, it does not mean that the given hybrid system has no invariants with the given degree bound $d$.*

## 4. Experiments

In this section, some examples are presented to illustrate our method for safety verification of hybrid systems.

**Example 1.** *[23, Example CLOCK] Consider a nonlinear continuous system*

$$\begin{bmatrix} \dot{x} \\ \dot{y} \end{bmatrix} = \begin{bmatrix} -\frac{11}{2}y + y^2 \\ 6x - x^2 \end{bmatrix},$$

*with location invariant* $\Psi = \{(x, y) \in \mathbb{R}^2 : 1 \le x \le 5 \wedge 1 \le y \le 5\}$. *The problem is to verify that all trajectories of the system starting from the initial set* $\Theta = \{(x, y) \in \mathbb{R}^2 : 4 \le x \le 4.5 \wedge y = 1\}$ *will never reach the unsafe set* $X_u = \{(x, y) \in \mathbb{R}^2 : 1 \le x \le 2 \wedge 2 \le y \le 3\}$. *The safety of the continuous system can be verified if we can find a polynomial* $\varphi(x, y)$ *which satisfies conditions in Theorem 5. We rewrite* $\Theta, X_u, \Psi$ *as follows*

$$\begin{aligned} \Theta &= \{(x, y) \in \mathbb{R}^2 : \theta_1(x, y) \ge 0 \wedge \theta_2(x, y) \ge 0 \wedge \theta_3(x, y) \ge 0\}, \\ \Psi &= \{(x, y) \in \mathbb{R}^2 : \psi_1(x, y) \ge 0 \wedge \psi_2(x, y) \ge 0\}, \\ X_u &= \{(x, y) \in \mathbb{R}^2 : \zeta_1(x, y) \ge 0 \wedge \zeta_2(x, y) \ge 0\}, \end{aligned}$$

*where*

$$\begin{aligned} &\theta_1(x, y) = (4 - x)(x - 4.5), \quad \theta_2(x, y) = y - 1, \quad \theta_3(x, y) = 1 - y, \\ &\psi_1(x, y) = (1 - x)(x - 5), \quad \psi_2(x, y) = (1 - y)(y - 5), \\ &\zeta_1(x, y) = (1 - x)(x - 2), \quad \zeta_2(x, y) = (2 - y)(y - 3). \end{aligned}$$

*Assuming* $\deg(\varphi(x, y)) = d$, *for* $d = 1, 2, \ldots$ *and the degree bound of all the involved SOSes in the program (5) is* $2e = 10$. *Then the SOS program (5) becomes*

$$\begin{aligned} \varphi(x, y) &= \sigma_0(x, y) + \sigma_1(x, y)\theta_1(x, y) + \sigma_2(x, y)\theta_2(x, y) + \sigma_3(x, y)\theta_3(x, y), \\ \dot{\varphi}(x, y) &= \phi_0(x, y) + \phi_1(x, y)\psi_1(x, y) + \phi_2(x, y)\psi_2(x, y) + \epsilon_1, \\ -\varphi(x, y) &= \mu_0(x, y) + \mu_1(x, y)\zeta_1(x, y) + \mu_2(x, y)\zeta_2(x, y) + \epsilon_2, \end{aligned}$$

*where* $\sigma_i(x, y), \phi_j(x, y), \mu_k(x, y) \in \Sigma_{2,2e}, \epsilon_1, \epsilon_2 \in \mathbb{R}_+$. *We apply the algorithm in Section 3.4, and increment d by 1 from* 1 *to* 10 *until a feasible solution of the SDP system is obtained. When* $d = 4$, *we obtain a feasible solution of the associated SDP system. Here we just list one approximate polynomial*

$$\begin{aligned} \varphi(x, y) = &-4.3296 - 1.2975x - 0.10418y + 0.92562x^2 + 0.18428xy \\ &+ 0.35738y^2 + \ldots + 0.94032 \times 10^{-6}x^4 + 0.17047 \times 10^{-5}y^4. \end{aligned}$$

*Let the tolerance $\tau = 10^{-2}$, and the bound of the common denominator of the polynomial coefficients vector be* 1000. *By use of the rational SOS recovery technique described in Section 3.3, we obtain all the corresponding polynomials with rational coefficients, for instance,*

$$\widetilde{\varphi}(x,y) = -\frac{4113}{950} - \frac{1233}{950}x - \frac{99}{950}y + \frac{879}{950}x^2 + \frac{34}{95}y^2 + \frac{7}{38}xy - \frac{6}{475}xy^2 - \frac{46}{475}x^3.$$

*Furthermore, a certificate of SOS representation shows $\widetilde{\varphi}(x,y)$ satisfies the conditions in Theorem 5* exactly. *Therefore, the safety of this continuous system is proved.*

**Example 2.** *[23, Example ECO] Consider a predator-prey hybrid system depicted in Figure 2, where*

$$f_1(\mathbf{x}) = f_2(\mathbf{x}) = \begin{pmatrix} -x_1 + x_1x_2 \\ x_2 - x_1x_2 \end{pmatrix}.$$

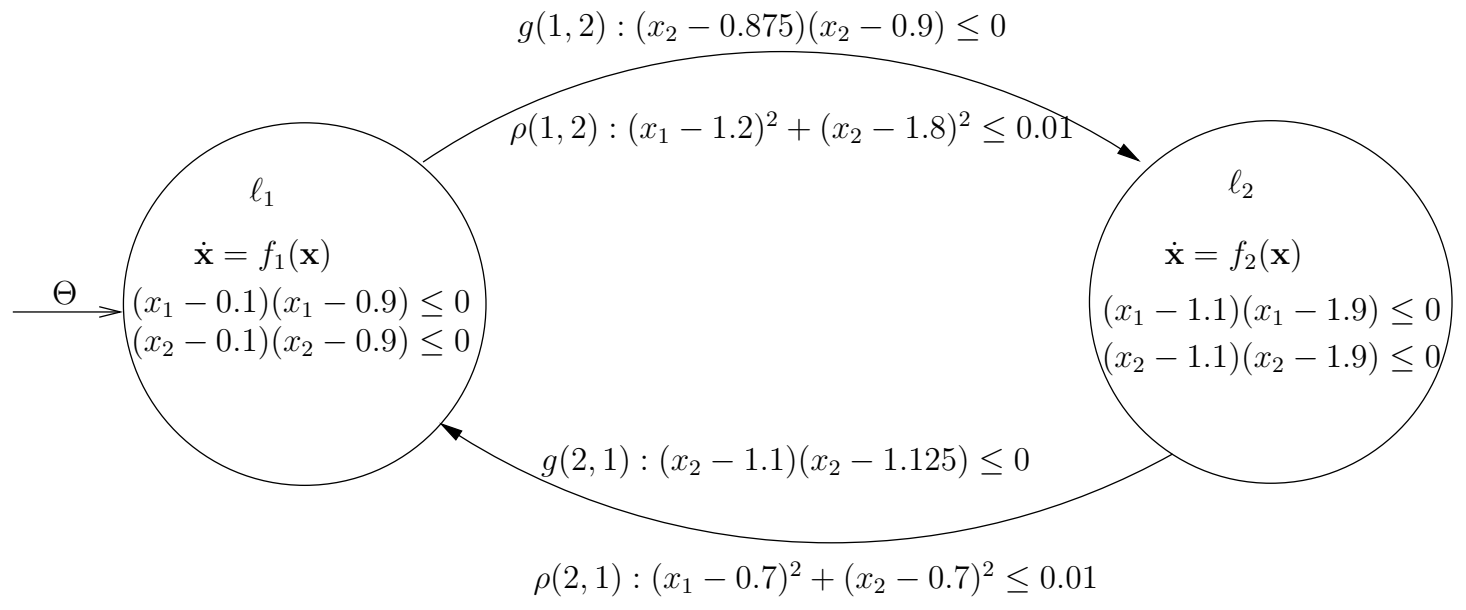

Figure 2: Hybrid system of example 2

*The system starts in location $\ell_1$, with an initial state in*

$$\Theta = \{(x_1, x_2) \in \mathbb{R}^2 : (x_1 - 0.8)^2 + (x_2 - 0.2)^2 \le 0.01\}.$$

*Our task is to verify the system never reach the states in*

$$X_u(\ell_1) = \{(x_1, x_2) \in \mathbb{R}^2 : 0.8 \le x_1 \le 0.9 \wedge 0.8 \le x_2 \le 0.9\}.$$

*To verify the safety of this system, we need find the corresponding invariant polynomials $\varphi_1(x_1, x_2)$ and $\varphi_2(x_1, x_2)$ at locations $\ell_1$ and $\ell_2$, respectively.*

*Similar to Example 1, we construct the associated SOS system, and find the feasible numerical solutions from SDP solver:*

$$\varphi_1 = 0.34871 - 0.45903x_1 + 0.018001x_2 + 0.2212x_1^2 - 0.45764x_1x_2 + 0..17991x_2^2,$$
$$\varphi_2 = 0.011167 + 1.2891x_1 + 0.56568x_2 + 0.88855x_1^2 - 0.56553x_1x_2 - 0..18386x_2^2.$$

*Let the tolerance* $\tau = 10^{-2}$*, and the bound of the common denominator of the polynomial coefficients vector be* 1000*. By use of the rational SOS recovery technique , we obtain all the corresponding polynomials with rational coefficients. The invariant polynomials with rational coefficients are*

$$\widetilde{\varphi}_1(x_1, x_2) = \frac{329}{944} - \frac{433}{944}x_1 + \frac{17}{944}x_2 - \frac{27}{59}x_1x_2 + \frac{209}{944}x_1^2 + \frac{85}{472}x_2^2,$$
$$\widetilde{\varphi}_2(x_1, x_2) = \frac{11}{944} + \frac{1217}{944}x_1 + \frac{267}{472}x_2 - \frac{267}{472}x_1x_2 + \frac{839}{944}x_1^2 - \frac{87}{472}x_2^2.$$

*Furthermore, all the remaining related polynomials in (5) can be written as SOSes of the polynomials, which means* $\widetilde{\varphi}_1$ *and* $\widetilde{\varphi}_2$ *satisfy all the conditions in Theorem 5. So the safety of hybrid system is proved.*

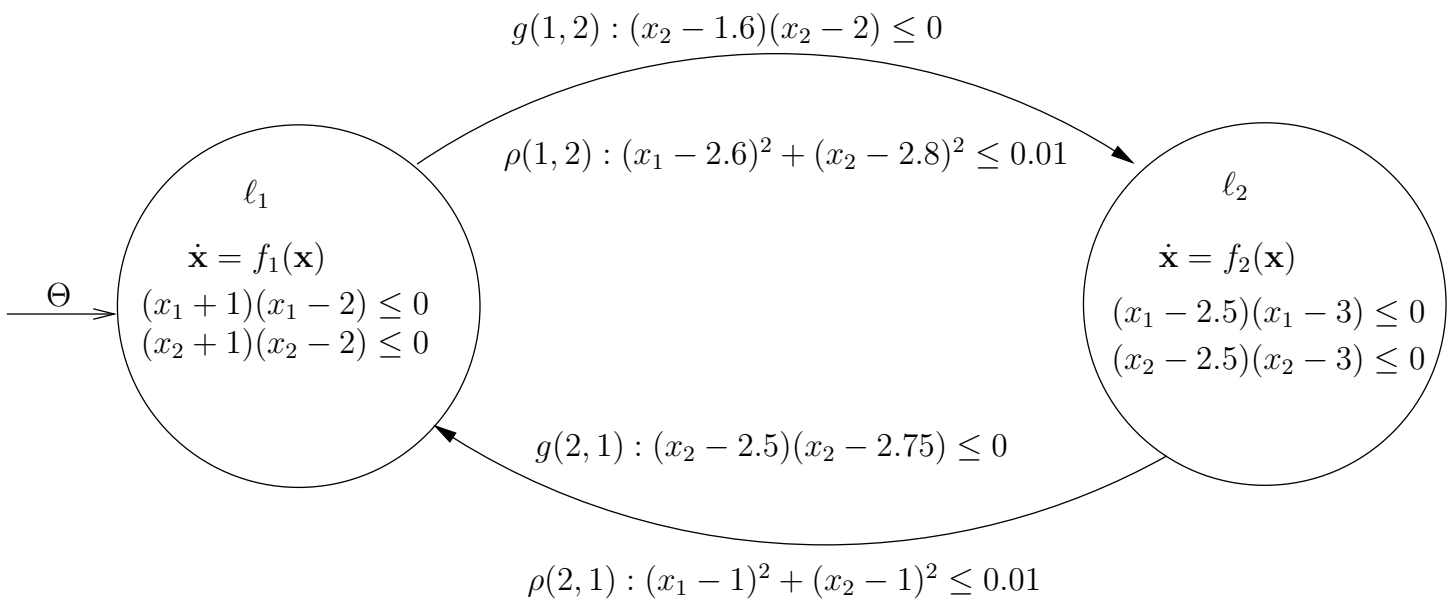

Figure 3: Hybrid system of example 3

**Example 3.** *Consider a hybrid system depicted in Figure 3, where*

$$f_1(\mathbf{x}) = \begin{pmatrix} 2x_1 + x_2 \\ x_1x_2 - x_2^2 - 1 \end{pmatrix}, \quad f_2(\mathbf{x}) = \begin{pmatrix} x_2 \\ -x_1 + x_2 \end{pmatrix}.$$

*The system starts in location* $\ell_1$*, with an initial state in*

$$\Theta = \{(x_1, x_2) \in \mathbb{R}^2 : (x_1 - 1.5)^2 + x_2^2 \leq 0.25\}.$$

*Our task is to verify the system never reach the states in*

$$X_u(\ell_1) = \{(x_1, x_2) \in \mathbb{R}^2 : (x_1 + 1)^2 + (x_2 + 1)^2 \leq 0.16\}.$$

*To prove the safety of the hybrid system, it suffices to find an inductive invariant polynomial $\varphi(x_1, x_2)$ which satisfies all the conditions in Theorem 6.*

*Using the same techniques illustrated in Examples 1 and 2, we obtain the inductive invariant polynomial with rational coefficients*

$$\widetilde{\varphi}(x_1, x_2) = -\frac{22}{49} + \frac{319}{931}x_1 - \frac{251}{931}x_2 + \frac{239}{931}x_1^2.$$

*Moreover, $\widetilde{\varphi}$ satisfies the conditions in Theorem 6 exactly. Therefore, the inductive invariant can guarantee the safety of the hybrid system. More details about the verification of conditions in Theorem 6, based on SOS representations of polynomials with rational coefficients can be found in Appendix.*

## 5. Conclusions

In this paper, we present a symbolic-numeric approach to compute inequality invariants for safety verification of hybrid systems. Employing SOS relaxation and rational vector recovery techniques, it can be guaranteed that an exact invariant, rather than a numerical one, can be obtained efficiently and practically. This approach avoids both the weakness of numerical approaches to verify safety of hybrid systems and the high complexity of symbolic invariant generation methods based on quantifier elimination.

## 6. Appendix

Solution to Example 3:

The initial state $\Theta$, the unsafe region $X_u(\ell_1)$, the state invariant $\Psi(\ell)$,

the guard condition $g(\ell, \ell')$ and the reset map $\rho(\ell, \ell')$ can be expressed as

$$
\begin{aligned}
&\Theta = \{(x_1, x_2) \in \mathbb{R}^2 : \theta_1(x_1, x_2) \geq 0\},\\
&\Psi(1) = \{(x_1, x_2) \in \mathbb{R}^2 : \psi_{1,1}(x_1, x_2) \geq 0 \wedge \psi_{1,2}(x_1, x_2) \geq 0\},\\
&\Psi(2) = \{(x_1, x_2) \in \mathbb{R}^2 : \psi_{2,1}(x_1, x_2) \geq 0 \wedge \psi_{2,2}(x_1, x_2) \geq 0\},\\
&g(1,2) = \{(x_1, x_2) \in \mathbb{R}^2 : g_{12}(x_1, x_2) \geq 0\},\\
&g(2,1) = \{(x_1, x_2) \in \mathbb{R}^2 : g_{21}(x_1, x_2) \geq 0\},\\
&\rho(1,2) = \{(x_1, x_2) \in \mathbb{R}^2 : \rho_{12}(x_1, x_2) \geq 0\},\\
&\rho(2,1) = \{(x_1, x_2) \in \mathbb{R}^2 : \rho_{21}(x_1, x_2) \geq 0\},\\
&X_u(1)(x_1, x_2) = \{(x_1, x_2) \in \mathbb{R}^2 : \zeta_1(x_1, x_2) \geq 0\},
\end{aligned}
$$

where

$$
\begin{aligned}
&\theta_1(x_1, x_2) = 0.25 - (x_1 - 1.5)^2 - x_2^2,\\
&\zeta_1(x_1, x_2) = 0.16 - (x_1 + 1)^2 - (x_2 + 1)^2,\\
&\psi_{1,1}(x_1, x_2) = (x_1 + 1)(2 - x_1),\ \psi_{1,2}(x_1, x_2) = (x_2 + 1)(2 - x_2),\\
&\psi_{2,1}(x_1, x_2) = (x_1 - 2.5)(3 - x_1),\ \psi_{2,2}(x_1, x_2) = (x_2 - 2.5)(3 - x_2),\\
&g_{12}(x_1, x_2) = (x_2 - 1.6)(2 - x_2),\ g_{21}(x_1, x_2) = (x_2 - 2.5)(2.75 - x_2),\\
&\rho_{12}(x_1, x_2) = 0.01 - (x_1 - 2.6)^2 - (x_2 - 2.8)^2,\\
&\rho_{21}(x_1, x_2) = 0.01 - (x_1 - 1)^2 - (x_2 - 1)^2.
\end{aligned}
$$

Let the tolerance $\tau = 10^{-2}$, and the bound of the common denominator of the polynomial coefficients vector be 1000, we can find that the inductive invariant polynomial $\widetilde{\varphi}(x_1, x_2)$ satisfies

$$
\begin{aligned}
&\widetilde{\varphi}(x_1, x_2) = \widetilde{\sigma}_0(x_1, x_2) + \widetilde{\sigma}_1(x_1, x_2)\theta_1(x_1, x_2),\\
&\widetilde{\varphi}(x_1, x_2) = \widetilde{\lambda}_{120}(x_1, x_2) + \widetilde{\lambda}_{121}(x_1, x_2)g_{12}(x_1, x_2) + \widetilde{\gamma}_{12}(x_1, x_2)\rho_{12}(x_1, x_2),\\
&\widetilde{\varphi}(x_1, x_2) = \widetilde{\lambda}_{210}(x_1, x_2) + \widetilde{\lambda}_{211}(x_1, x_2)g_{21}(x_1, x_2) + \widetilde{\gamma}_{21}(x_1, x_2)\rho_{21}(x_1, x_2),\\
&\dot{\widetilde{\varphi}}(x_1, x_2) = \widetilde{\phi}_{10}(x_1, x_2) + \widetilde{\phi}_{11}(x_1, x_2)\psi_{11}(x_1, x_2) + \widetilde{\phi}_{12}(x_1, x_2)\psi_{12}(x_1, x_2) + \epsilon_1,\\
&\dot{\widetilde{\varphi}}(x_1, x_2) = \widetilde{\phi}_{20}(x_1, x_2) + \widetilde{\phi}_{21}(x_1, x_2)\psi_{21}(x_1, x_2) + \widetilde{\phi}_{22}(x_1, x_2)\psi_{22}(x_1, x_2) + \epsilon_2,\\
&-\widetilde{\varphi}(x_1, x_2) = \widetilde{\mu}_0(x_1, x_2) + \widetilde{\mu}_1(x_1, x_2)\zeta_1(x_1, x_2) + \widetilde{\epsilon},
\end{aligned}
$$

where

$$
\begin{aligned}
&\widetilde{\varphi}(x_1,x_2)=-\tfrac{22}{49}+\tfrac{319}{931}x_1-\tfrac{251}{931}x_2+\tfrac{239}{931}x_1^2,\\
&\widetilde{\sigma}_0(x_1,x_2)=\tfrac{838}{931}-\tfrac{1565}{931}x_1-\tfrac{251}{931}x_2+\tfrac{867}{931}x_1^2+\tfrac{628}{931}x_2^2,\quad \widetilde{\sigma}_1=\tfrac{628}{931},\\
&\widetilde{\lambda}_{120}(x_1,x_2)=\tfrac{8403}{1900}-\tfrac{4463}{4655}x_1-\tfrac{14169}{4655}x_2+\tfrac{472}{931}x_1^2+\tfrac{12}{19}x_2^2,\\
&\widetilde{\lambda}_{121}=\tfrac{355}{931},\quad \widetilde{\gamma}_{12}=\tfrac{233}{931},\\
&\widetilde{\lambda}_{210}(x_1,x_2)=\tfrac{133187}{37240}-\tfrac{1271}{931}x_1-\tfrac{1997}{532}x_2+\tfrac{1034}{931}x_1^2+\tfrac{1110}{931}x_2^2,\\
&\widetilde{\lambda}_{211}=\tfrac{45}{133},\quad \widetilde{\gamma}_{21}=\tfrac{795}{931},\\
&\widetilde{\phi}_{10}(x_1,x_2)=\tfrac{153}{931}+\tfrac{89}{133}x_1+\tfrac{298}{931}x_2+\tfrac{227}{931}x_1x_2+\tfrac{971}{931}x_1^2+\tfrac{272}{931}x_2^2,\\
&\widetilde{\phi}_{11}=\tfrac{15}{931},\quad \widetilde{\phi}_{12}=\tfrac{3}{133},\quad \widetilde{\epsilon}_1=\tfrac{26}{931},\\
&\widetilde{\phi}_{20}(x_1,x_2)=\tfrac{3751}{931}-\tfrac{3337}{1862}x_1-\tfrac{3373}{1862}x_2+\tfrac{349}{931}x_1^2+\tfrac{478}{931}x_1x_2+\tfrac{319}{931}x_2^2,\\
&\widetilde{\phi}_{21}=\tfrac{349}{931},\quad \widetilde{\phi}_{22}=\tfrac{319}{931},\quad \widetilde{\epsilon}_2=\tfrac{1259}{931},\\
&\widetilde{\mu}_0(x_1,x_2)=\tfrac{672}{485}-\tfrac{18}{97}x_1+\tfrac{137}{97}x_2+\tfrac{9}{97}x_1x_2+\tfrac{66}{97}x_1^2+\tfrac{92}{97}x_2^2,\\
&\widetilde{\mu}_1=\tfrac{564}{931},\quad \widetilde{\epsilon}=\tfrac{58}{931}.
\end{aligned}
$$

The exact SOS representations of above polynomials are as follows:

$$
\begin{aligned}
\widetilde{\sigma}_0(x_1,x_2)&=\tfrac{931}{838}h_{11}^2+\tfrac{3120712}{2042055}h_{12}^2+\tfrac{380230641}{46469677}h_{13}^2,\\
\widetilde{\lambda}_{120}(x_1,x_2)&=\tfrac{1900}{8403}h_{21}^2+\tfrac{127778819}{13782225}h_{22}^2+\tfrac{12831251475}{2601209876}h_{23}^2,\\
\widetilde{\lambda}_{210}(x_1,x_2)&=\tfrac{37240}{133187}h_{31}^2+\tfrac{991976776}{205638355}h_{32}^2+\tfrac{38289861701}{13835779654}h_{33}^2,\\
\widetilde{\phi}_{10}(x,y)&=\tfrac{931}{153}h_{41}^2+\tfrac{142443}{19415}h_{42}^2+\tfrac{72301460}{4096293}h_{43}^2,\\
\widetilde{\phi}_{20}(x,y)&=\tfrac{931}{3751}h_{51}^2+\tfrac{55874896}{7767975}h_{52}^2+\tfrac{7231984725}{1110827906}h_{53}^2,\\
\widetilde{\mu}_0(x_1,x_2)&=\tfrac{3325}{4992}h_{61}^2+\tfrac{978432}{235283}h_{62}^2+\tfrac{29133446909}{941925575}h_{63}^2,
\end{aligned}
$$

where

$$
\begin{aligned}
&h_{11}=\tfrac{838}{931}-\tfrac{251}{1862}x_2-\tfrac{1565}{1862}x_1,\ h_{12}=\tfrac{2042055}{3120712}x_2-\tfrac{392815}{3120712}x_1,\ h_{13}=\tfrac{46469677}{380230641}x_1.\\
&h_{21}=\tfrac{8403}{1900}-\tfrac{14169}{9310}x_2-\tfrac{4463}{9310}x_1,\ h_{22}=\tfrac{13782225}{127778819}x_2-\tfrac{21078749}{127778819}x_1,\ h_{23}=\tfrac{2601209876}{12831251475}x_1.\\
&h_{31}=\tfrac{133187}{37240}-\tfrac{1997}{1064}x_2-\tfrac{1271}{1862}x_1,\ h_{32}=\tfrac{205638355}{991976776}x_2-\tfrac{12690935}{35427742}x_1,\ h_{33}=\tfrac{13835779654}{38289861701}x_1.\\
&h_{41}=\tfrac{153}{931}+\tfrac{149}{931}x_2+\tfrac{89}{266}x_1,\ h_{42}=\tfrac{19415}{142443}x_2-\tfrac{29048}{142443}x_1,\ h_{43}=\tfrac{4096293}{72301460}x_1.\\
&h_{51}=\tfrac{3751}{931}-\tfrac{3373}{3724}x_2-\tfrac{3337}{3724}x_1,\ h_{52}=\tfrac{7767975}{55874896}x_2+\tfrac{3088123}{55874896}x_1,\ h_{53}=\tfrac{1110827906}{7231984725}x_1.\\
&h_{61}=\tfrac{4992}{3325}+\tfrac{197}{266}x_2+\tfrac{809}{1862}x_1,\ h_{62}=\tfrac{235283}{978432}x_2-\tfrac{3984325}{18590208}x_1,\ h_{63}=\tfrac{941925575}{29133446909}x_1.
\end{aligned}
$$